\documentclass[final]{IEEEtran}
\pdfoutput=1

\usepackage{amsmath}
\usepackage{amssymb}
\usepackage{graphicx}
\usepackage{subcaption}
\usepackage{xcolor}
\usepackage{multirow}
\usepackage{fancyhdr}

\usepackage{url}
\usepackage[sort]{cite}

\usepackage{hyperref}
\newcommand{\diag}{\mathop{\mathrm{diag}}}

\newcommand{\Var}{\operatorname{Var}}

\title{BER Performance Analysis of Coarse Quantized Uplink Massive MIMO}
\author{Azad~Azizzadeh,~%\IEEEmembership{Member,~IEEE,}
	Reza~Mohammadkhani,~%~\IEEEmembership{Fellow,~OSAMember,~IEEE,}
	Seyed~Vahab~Al-Din~Makki,~%~\IEEEmembership{Member,~IEEE}
	and~Emil~Bj\"ornson%~\IEEEmembership{Member,~IEEE}% <--this % stops a space
	\thanks{Corresponding author: \emph{R. Mohammadkhani}.}%
	\thanks{A. Azizzadeh and S. V. Makki are with the Department of Electrical Engineering, Razi University, Kermanshah, Iran
		(e-mail: azizzadeh.azad@razi.ac.ir, v.makki@razi.ac.ir).}% <--this % stops a space
	\thanks{R. Mohammadkhani is with the Department of Electrical Engineering, University of Kurdistan, Sanandaj, Iran (e-mail: r.mohammadkhani@uok.ac.ir).}%
	\thanks{E. Bj\"ornson is with the Department of Electrical Engineering (ISY), Linkoping University, Sweden (email: emil.bjornson@liu.se).}% <--this % stops a space
}

\begin{document}

\maketitle
\fancypagestyle{firststyle}
{
	\fancyhf{}% Clear header/footer
	\fancyhead[L]{This work has been submitted to the IEEE Journals for possible publication. Copyright may be transferred without notice, after which this version may no longer be accessible.}
	%	\fancyfoot[C]{\footnotesize Page \thepage\ of \pageref{LastPage}}
}
\thispagestyle{firststyle}

\begin{abstract}
Having lower quantization resolution, has been introduced in the literature, as a solution to reduce the power consumption of massive MIMO and millimeter wave MIMO systems. In this paper, we analyze bit error rate (BER) performance of quantized uplink massive MIMO  employing a few-bit resolution ADCs.
Considering Zero-Forcing (ZF) detection, 
we derive a closed-form quantized signal-to-interference-plus-noise ratio (SINR) to achieve an analytical BER approximation for coarse quantized M-QAM massive MIMO systems, by using a linear quantization model. The proposed expression is a function of quantization resolution in bits. 
We further numerically investigate the effects of different quantization levels, from 1-bit to 4-bits, on the BER of three modulation types of QPSK, 16-QAM, and 64-QAM. Uniform and non-uniform quantizers are employed in our simulation.

Monte Carlo simulation results reveal that our approximate formula gives a tight upper bound for the BER performance of $b$-bit resolution quantized systems using non-uniform quantizers, whereas the use of uniform quantizers cause a lower performance for the same systems. 
We also found  a small BER performance degradation in coarse quantized systems, for example 2-3 bits QPSK and 3-4 bits 16-QAM, compared to the full-precision (unquantized) case. 
However, this performance degradation can be compensated by increasing the number of antennas at the BS. 

\end{abstract}


%------------------------------------------------------ Introduction
\section{Introduction}
\label{sec:Intro}
\IEEEPARstart{M}{assive} MIMO technology as a result of rethinking the concept of MIMO wireless communications, enables each base station (BS) to communicate with tens of users at the same time and frequency, by increasing the number of antennas at the BS \cite{larsson2014massive}.
Furthermore, this technology reduces the effect of additive thermal noise for the uplink by averaging over a large array at the BS, and allows the use of simple linear processing techniques \cite{rusek2013scaling}.

However, massive MIMO systems, having hundreds of antennas and the same number of radio frequency (RF) chains at the BS, are facing high power consumption and hardware complexities. Among hardware components of each RF chain, analog-to-digital-converter (ADC) has attracted the most interest. It stands to reason  that power consumption of an ADC is growing exponentially by increasing the quantization resolution, and linearly by an increase in sampling rate or bandwidth \cite{walden1999analog,singh2009multi}.
Moreover, there is a limit of ({\small\textsf{sampling rate} $\times$ \textsf{bit-resolution}}) for ADCs \cite{walden1999analog}. 
Therefore, several studies have investigated the use of low resolution quantization (in bits) for massive MIMO \cite{fan2015uplink,wang2014multiuser,zhang2016spectral,wang2014convex,risi2014massive,zhang2015mpdq,desset2015validation} and millimeter wave MIMO systems \cite{orhan2015low,mo2014high,heath2016overview}.

Reducing the bit-resolution of ADCs, results in the reduction of power consumption not only for the ADCs, but also for the baseband circuits connected to ADC/DAC \cite{heath2016overview}. 
However, by the use of few-bit resolution and especially the ultimate coarse quantization level of 1-bit ADCs, we face several challenges. Channel estimation algorithms \cite{wen2016bayes,choi2016near}, the way we use channel-state-information (CSI) for precoding \cite{saxena2017analysis,jacobsson2017quantized,jedda2016minimum}, detection techniques \cite{mezghani2007modified,mezghani2012iterative,wang2014multiuser,zhang2015mpdq} and other signal processing algorithms are different.

Several studies have investigated the effects of quantization bit-resolution on achievable rates \cite{orhan2015low,fan2015uplink,mollen2017achievable} and energy efficiency (ratio of data rate to the power consumption)\cite{sarajlic2017low} recently, and some closed-form expressions are proposed for the achievable rate of quantized massive MIMO systems \cite{mollen2017uplink,fan2015uplink}. 

However, up to our knowledge, there is no theoretical closed-form expression in the literature for the BER performance of low-resolution (in bit) quantized massive MIMO systems. 
There are some simulation results that investigate the effects of low-resolution (in bits) quantization on the BER performance for downlink \cite{desset2015validation}, and uplink \cite{azizzadeh2017ber} massive MIMO systems. However, available analytical studies are limited to some special cases. For example, \cite{risi2014massive,saxena2017analysis} evaluate the Symbol-Error-Rate (SER) for 1-bit QPSK, at uplink and downlink, respectively. In addition, \cite{jacobsson2017quantized} studies the BER of quantized massive MIMO systems with different bit-resolutions, but only for QPSK modulation at downlink in order to rather design a precoder at the BS.
%Although, some simulation results are presented in \cite{desset2015validation,azizzadeh2017ber}, available analytical studies are limited to the special case of 1-bit ADCs \cite{saxena2017analysis} or DACs \cite{risi2014massive}.
%y is found, except \cite{saxena2017analysis} that presents a closed-form SER for the special case of 1-bit DACs in order to rather design a linear ZF precoder at the BS. 

In this paper, we study the BER performance of uplink massive MIMO systems with different coarse quantization levels of $b$-bit resolution ADCs. We present an approximate BER expression for M-QAM modulations, assuming ZF detection at the BS, using the liner quantization model. 
We extend our preliminary simulation results in \cite{azizzadeh2017ber} that uses uniform quantizers, to the case of having both uniform and non-uniform quantizers.
Our contributions are listed as follows:
\begin{itemize}
\item Obtaining the ZF detection matrix for the $b$-bit resolution quantized system, using the liner quantization model,
\item Deriving a quantized signal-to-interference-plus-noise ratio (SINR) that leads to a closed-form BER expression for M-QAM quantized massive MIMO systems,
\item Evaluating asymptotic BER performance of quantized systems
\item Simulating the BER of quantized massive MIMO systems with different $b$-bit resolutions from $b=1$ to 4 for three modulation types of QPSK, 16-QAM, and 64-QAM, using both uniform and non-uniform quantizers.
\item Simulating the BER degradation to find the optimum $b$-bit quantization resolution for each one of the above modulations. 
\end{itemize}

The rest of this paper is organized as follows. 
Section \ref{sec:model} presents the system model, and reviews a linear quantizer model based on Bussgang decomposition theory for Gaussian distributed signals, that would be simplified to a simple model called additive quantization noise model (AQNM), when we have equal $b$-bit quantizers for all antennas.
Then, considering ZF detection for massive MIMO systems in Section \ref{sec:ZF_Q_MassiveMIMO}, we derive a ZF detection matrix for $b$-bit resolution quantized systems, using the linear quantizer model.
Next, in Section \ref{sec:BER_QuantizedMassiveMIMO}, a BER expression of M-QAM MIMO using ZF detectors is used and extended to apply in quantized massive MIMO systems, followed by asymptotic BER behavior analysis of such systems. Section \ref{sec:Simulations} provides the numerical BER results employing ($b=1$ to 4)-bit resolution ADCs for three modulation types of QPSK, 16-QAM, and 64-QAM, using both uniform and non-uniform quantizers. At the end, we conclude the paper in Section \ref{sec:conclusion}.

%------------------------------------------------------ System Model
\section{System Model}
\label{sec:model}
An uplink massive MIMO system with one base station (BS) having $N$ antennas, and serving $K$ single-antenna user equipments is considered. We assume users transmit a symbol vector $\mathbf x \in \mathbb{C}^{K\times 1}$ where each symbol $x_k$ has a constellation size $M$. The received symbol vector $\mathbf y \in \mathbb{C}^{N\times 1}$ at the BS, is given by
\begin{equation}
\mathbf y =\mathbf  H \mathbf x +\mathbf n=\sum_{k=1}^K \mathbf h_k x_k+\mathbf n
\label{eq:y_Hx+n}
\end{equation}
where $\mathbf h_k \in \mathbb{C}^{N\times 1}$ is the channel vector between the BS and the $k$th user, $\mathbf H\overset{\Delta}{=} [\mathbf h_1,\mathbf h_2,...,\mathbf h_K]\in \mathbb{C}^{N\times K}$ denotes the channel matrix, and $\mathbf n \sim CN(0,\sigma^2_n\mathbf I_N)$  is the additive white Gaussian noise vector. The channel state information (CSI) is unknown to the users (transmit side), therefore we assume the same symbol energy per user, and $\mathbf R_{\mathbf x}=E\{\mathbf x\mathbf x^H\}=\sigma^2_x\mathbf I_K$. We further assume that entries of $\mathbf H$ and $\mathbf n$ are independent.

\begin{figure}[t]
	\centering
	\includegraphics[width=0.99\linewidth]{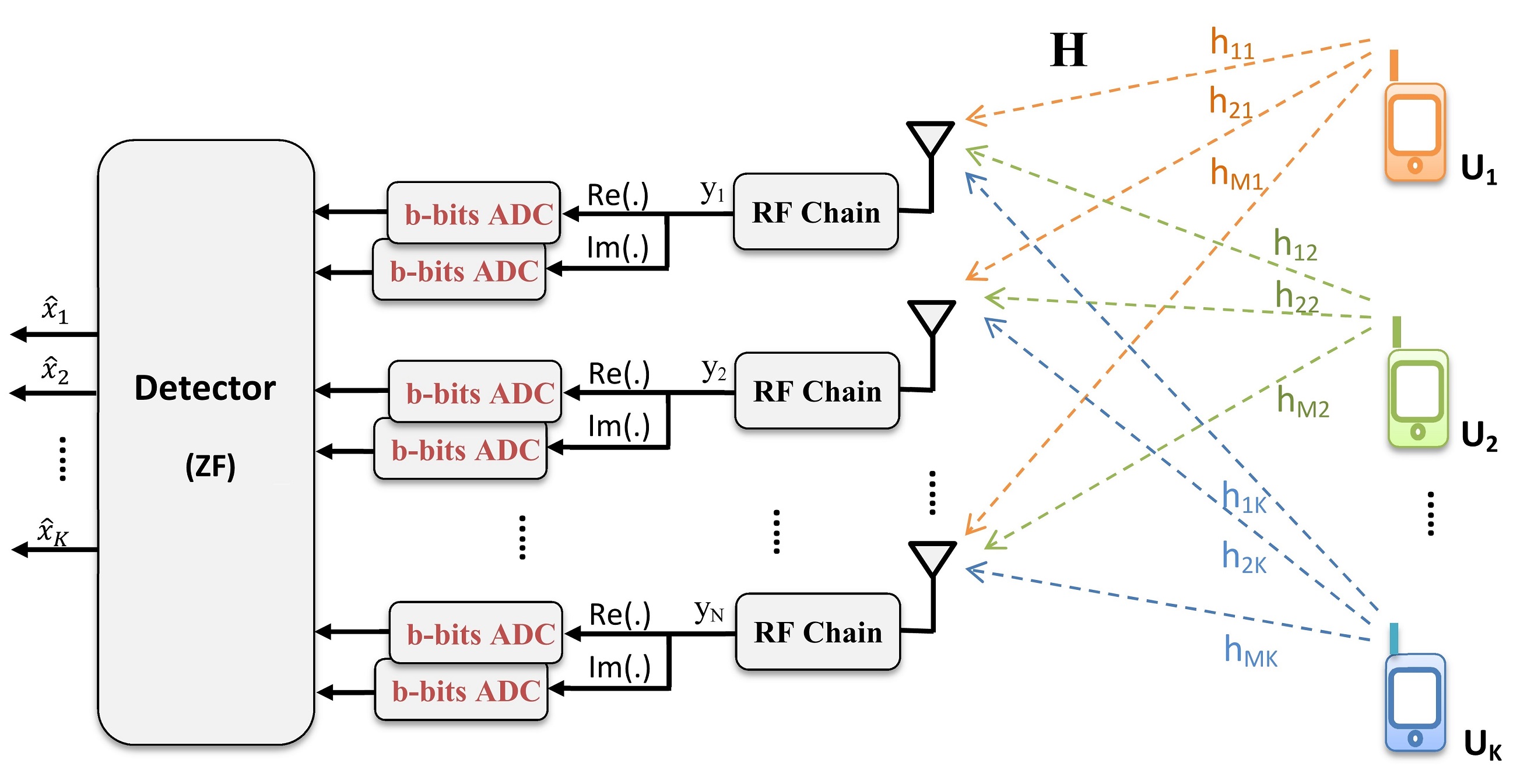}
	\caption{An uplink quantized massive MIMO system}
	\label{fig:ADC_massive}
\end{figure}

As illustrated in Fig.\ref{fig:ADC_massive}, the real and imaginary parts of the complex received signal at each antenna, are quantized separately by $b$-bit resolution ADCs. The resulting quantized signal vector is defined as
\begin{align}
 \mathbf y_q &=Q(\mathbf y)=Q\left(\mathbf y_R\right)+j Q(\mathbf y_I)\nonumber\\
             &= \mathbf y + \mathbf e
 \label{eq:yq_Q}
\end{align}
where $Q(\cdot)$ represents the quantization function, $\mathbf e$ is the vector of quantization error, and $\mathbf y_R$ and $\mathbf y_I$ are the real and imaginary parts of $\mathbf y$, respectively.

\begin{table}
	\renewcommand{\arraystretch}{1.5}
	\centering
	\caption{values of $\alpha$ and $\rho$   for $b$-bit resolution ADCs}
	\label{tab:rho}
	\begin{tabular}{|c|c|c|c|c|c|}\hline
		$b$	    & 1		 & 	2	  & 	3	& 	4	  & 5	\\ \hline\hline
		$\rho$	& 0.3634 & 0.1175 & 0.03454 & 0.009497& 0.002499\\ \hline
		$\alpha$& 0.6366 & 0.8825 & 0.96546 & 0.990503& 0.997501\\ \hline
	\end{tabular}
\end{table}

\subsection{Linear Quantizer Model}

Assuming a Gaussian input vector $\mathbf x$ in (\ref{eq:y_Hx+n}), for each realization of the channel matrix $\mathbf H$, the output $\mathbf y$ would also be Gaussian distributed. Therefore, according to the Bussgang theorem \cite{bussgang1952crosscorrelation,mezghani2012iterative}, the output of the non-linear quantizer $\mathbf y_q=Q(\mathbf y)$ can be decomposed into a desired signal part and an uncorrelated distortion, as follows
\begin{equation}
\mathbf y_q=\mathbf B\mathbf  y + \mathbf  n_q, 
\label{eq:yq_bussgang}
\end{equation}
where the quantization noise vector $\mathbf n_q$ is  uncorrelated with $\mathbf y$, and $\mathbf B$ is a linear matrix operator chosen to minimize the power of the quantization noise $\mathbf n_q$, and is given by\cite{mezghani2012iterative,li2017channel}
\begin{equation}
\mathbf B  =\mathbb{E} \left[\mathbf y_q \mathbf y^H\right] \mathbb{E}\left[\mathbf y \mathbf y^H\right]^{-1} 
           = \mathbf R_{\mathbf y_q\mathbf y}\mathbf R_{\mathbf y\mathbf y}^{-1}.
\end{equation}

Let assume $y_i$ as the received signal of the $i$th antenna, quantized by two separate $b$-bit quantizers for the real and imaginary parts, $y_{i,R}$ and $y_{i,I}$, respectively. % of $y_i=y_{i,R}+j y_{i,I}$.  
Therefore, we have  $y_q^{i,R}=Q(y_{i,R})$ and $y_q^{i,I}=Q(y_{i,I})$, and distortion factor for each quantizer can be expressed as \cite{mezghani2012iterative}
\begin{equation}
	\rho^{i,c}=\frac{\Var[e_{i,c}]}{\Var[y_{i,c}]}
\end{equation} 
for $i=1,2,\dots,N$ and $c\in \{R,I\}$, where $\Var[y_{i,c}]$ is the variance of the input $y_{i,c}$, and $\Var[e_{i,c}]$ is the variance of the quantizer error $e_{i,c}=y_q^{i,c}-y_{i,c}$. It is worth noting that the distortion factor $\rho$ of each quantizer is equal to the inverse of the signal-to-quantization-noise ratio (SQNR).

Assuming the same $b$-bit resolution ADCs for all received signals (and for both real and imaginary parts of each one), in a MIMO system illustrated in Fig. \ref{fig:ADC_massive}, and considering Gaussian distributed $\mathbf x$ and $\mathbf y$, the distribution factor of all quantizers would be equal, i.e. $\rho^{i,c}=\rho$ for all $i$ and $c\in\{R,I\}$. 
Therefore, for each realization of the channel matrix $\mathbf H$, the correlation matrix of $\mathbf n_q$ can be expressed as \cite{mezghani2012iterative,bai2013optimization}
\begin{align}
\mathbf R_{\mathbf n_q\mathbf n_q}  
&=\mathbb{E} \left[(\mathbf y_q- \mathbf B\mathbf y)(\mathbf y_q- \mathbf B\mathbf y)^H\right] \nonumber\\ 
&= \mathbf R_{\mathbf y_q\mathbf y_q}-\mathbf R_{\mathbf y_q\mathbf y}\mathbf R_{\mathbf y\mathbf y}^{-1}\mathbf R_{\mathbf y\mathbf y_q} \nonumber\\ 
&=\rho(1-\rho) \diag\left(\mathbf R_{\mathbf y\mathbf y}\right)
\label{eq:R_nq}
\end{align}
where $\rho$ is a scalar depending on the number of quantization bit-resolution\footnote{However, it should be replaced by a diagonal matrix $\boldsymbol{\rho}\in\mathbb{R}^{N\times N}$ if we use different bit-resolution for the available quantizers \cite{bai2013optimization}.}. 
For such assumptions, the linear matrix $\mathbf B$ can also be simplified as
\begin{equation}
	\mathbf B=(1-\rho)\mathbf I_N=\alpha \mathbf I_N.
	\label{eq:B_rho}
\end{equation}
Table \ref{tab:rho} shows the values of $\rho$ and $\alpha$ for $b$-bit quantization resolution ($b\le 5$). %, when we have an optimal minimum mean squared error (MMSE) quantizer with a standard Gaussian input signal \cite{max1960quantizing}. 
For higher bit-resolution of $b>5$, an approximate of $\rho = \frac{\pi\sqrt{3}}{2} 2^{-2b}$ can be applied \cite{bai2015energy}.
%
%
%
%Referring back to (\ref{eq:y_Hx+n}) and (\ref{eq:yq_Q}), we can extend (\ref{eq:aqnm_scalar}) to a vector form of
%Assuming the same $b$-bit resolution ADCs for all received signals (and for both real and imaginary parts of each one), in a MIMO system illustrated in Fig. \ref{fig:ADC_massive}, we can extend (\ref{eq:aqnm_scalar}) to a vector form of
%There is another model in the literature called \emph{Additive Quantization Noise Model} (AQNM) that approximate the quantizer by a linear gain with additive white noise, given by \cite{fletcher2007robust,orhan2015low,fan2015uplink}. 
Substituting (\ref{eq:B_rho}) into (\ref{eq:yq_bussgang}), we have
\begin{equation}
	\mathbf y_q=(1-\rho)\mathbf  y + \mathbf  n_q =\alpha\mathbf  y + \mathbf  n_q.
	\label{eq:yq_aqnm}
\end{equation}
The above form of the quantizer model is called \emph{Additive Quantization Noise Model} (AQNM) in the current studies \cite{fletcher2007robust,orhan2015low,fan2015uplink}. As we see, the linear quantizer model resulted from Bussgang decomposition theory, considering equal $b$-bit quantizers for all antennas, will be equal to the AQNM model.

%Given the channel matrix $\mathbf H$, and considering our assumption for $\mathbf R_{\mathbf x}$, the covariance matrix of the noise vector $\mathbf n_q$ can be expressed as \cite{fan2015uplink}
%\begin{align}
%\mathbf R_{\mathbf n_q} & =\mathbb{E} \lbrace \mathbf n_q \mathbf n_q^H \vert \mathbf H\rbrace=
%\alpha (1-\alpha)\mathbb{E} \lbrace \mathbf y \mathbf y^H \vert \mathbf H\rbrace \nonumber\\ 
%& = \alpha (1-\alpha)\diag(\mathbf H \mathbf R_{\mathbf x} \mathbf H^H+\sigma_n^2\mathbf I_N)\nonumber\\
%& = \alpha (1-\alpha)\diag(\sigma_x^2\mathbf H \mathbf H^H+\sigma_n^2\mathbf I_N)
%\end{align}
%where $\mathbf I_N$ is an $N\times N$ identity matrix.

%------------------------------------------------------ ZF for Quantized Massive MIMO
\section{ZF Detection of Quantized Massive MIMO}
\label{sec:ZF_Q_MassiveMIMO}
In this section, we begin with ZF detection for unquantized massive MIMO systems, and then, we extend it to the $b$-bit resolution quantized systems by using the linear quantizer model.

\subsection{ZF Detection}
 It has been known that linear detectors (such as zero forcing (ZF) and minimum mean squared error (MMSE) detectors) have lower computational complexity  compared to optimal detectors, at the expense of achieving suboptimal performance \cite{brown2012practical,bai2012low}.
However, for a massive MIMO system where $N\gg K\gg 1$, linear detectors perform very close to optimal detectors \cite{ngo2013energy,ma2014massive}. 

In this article, we employ ZF detector for an uplink massive MIMO scenario. Having the channel state information at the BS (receive side), we multiply a $K\times N$ detection matrix $\mathbf A^H$ by the received vector $\mathbf y$, to have an estimate of the transmit symbol vector $\mathbf x$ as follows
\begin{equation}
\hat{\mathbf x} =\mathbf A^H \mathbf y=\mathbf A^H\mathbf H\mathbf x+\mathbf A^H\mathbf n.
\label{eq:x_hat}
\end{equation}
Following the fact that noise is ignored in ZF detection, the detection matrix is found by solving 
\begin{equation}
\mathbf A^H\mathbf H\mathbf x=\mathbf x.
\end{equation}
Since $\mathbf H$ is a non-square matrix in general, we may express the solution as\cite{ma2014massive}
\begin{equation}
\mathbf A = \mathbf H (\mathbf H^H\mathbf H)^{-1}.
\label{eq:A_ZF}
\end{equation}
Substituting (\ref{eq:A_ZF}) into (\ref{eq:x_hat}), an estimate of the transmit symbol vector is given by
\begin{equation}
\hat{\mathbf x} =\mathbf x+(\mathbf H^H\mathbf H)^{-1} \mathbf H^H \mathbf n
\label{eq:x_hat_ZF} 
\end{equation}
and by doing some maths, we can express the signal-to-interference-noise-ratio (SINR) of the $k$th user as
\begin{equation}
\gamma_k = \frac{\sigma_x^2}{\sigma_n^2[(\mathbf H^H\mathbf H)^{-1}]_{kk}}
 = \gamma_0\frac{1}{[(\mathbf H^H\mathbf H)^{-1}]_{kk}}
\label{eq:gamma_k}
\end{equation}
where $\gamma_0={\sigma_x^2}/{\sigma_n^2}$ is the SNR for AWGN channel model (in other words, when $\mathbf H$ is an identity matrix), and $[\cdot]_{kk}$ denotes the $k$th diagonal element. For an independent and identically distributed (i.i.d) Rayleigh flat-fading channel matrix $\mathbf H$, the article \cite{winters1994impact} shows that $\chi^2_d={1}/{[(\mathbf H^H\mathbf H)^{-1}]_{kk}}$ is a chi-squared random variable with $d=2(N-K+1)$ degrees of freedom. Since the SINR distribution of symbol streams for all users are assumed to be equal, i.e. uniform power allocation for the case of no CSI at the transmit side, we simply neglect the subscript $k$ and consider the SINR of each symbol stream as $\gamma=\gamma_0 \chi^2_d$.

\subsection{ZF Detection of Quantized Massive MIMO}
In order to investigate the effects of low resolution quantization on the BER performance of massive MIMO systems, we substitute the linear quantizer model of (\ref{eq:yq_aqnm}) into (\ref{eq:x_hat}) as follows
\begin{align}
\hat{\mathbf x} 
&=\mathbf A^H \mathbf y_q\approx\mathbf A^H(\alpha\mathbf y+\mathbf n_q)\nonumber\\
&=\alpha\mathbf A^H\mathbf H\mathbf x+\underbrace{\mathbf A^H(\alpha\mathbf n+\mathbf n_q)}_{noise}.                              
\label{eq:xhat_A_yq}
\end{align}
Therefore, we have to solve
\begin{equation}
\alpha\mathbf A^H\mathbf H\mathbf x = \mathbf x,
\end{equation}
to find the ZF detection matrix for a quantized system, using linear quantization model. It is given by
\begin{equation}
\mathbf A_q = \frac{1}{\alpha}\mathbf H (\mathbf H^H\mathbf H)^{-1}.
\label{eq:A_ZF_q}
\end{equation}
Substituting (\ref{eq:A_ZF_q}) into (\ref{eq:xhat_A_yq}), we may write the transmit vector estimate as
\begin{align}
\hat{\mathbf x} 
&=\mathbf A_q^H \mathbf y_q\approx\mathbf A_q^H(\alpha\mathbf y+\mathbf n_q)\nonumber\\
&=\frac{1}{\alpha}(\mathbf H^H\mathbf H)^{-1}\mathbf H^H(\alpha\mathbf H\mathbf x+\alpha\mathbf n+\mathbf n_q) \nonumber \\
&=\mathbf x+(\mathbf H^H\mathbf H)^{-1}\mathbf H^H(\mathbf n+\frac{1}{\alpha}\mathbf n_q) \nonumber \\
&=\mathbf x+\mathbf n_e 
\label{eq:xhat_Aq_yq}
\end{align}
we define $\mathbf n_e = (\mathbf H^H\mathbf H)^{-1}\mathbf H^H(\mathbf n+\frac{1}{\alpha}\mathbf n_q)$ as the \emph{effective noise}, consisting of an additive white Gaussian noise (AWGN) and the quantization noise. 
In order to determine the SINR of user $k$, we need to calculate the covariance matrix of the effective noise
%Now, we determine SINR of the estimated symbol streams as
%\begin{equation}
%\text{SINR} = \frac{\mathbb{E}\lbrace \mathbf x \mathbf x^H\rbrace}{\mathbb{E}\lbrace \mathbf n_e \mathbf n_e^H\rbrace}= \frac{\sigma_x^2\mathbf I_K}{\mathbb{E}\lbrace \mathbf n_e \mathbf n_e^H\rbrace}.
%\label{eq:SINR_xhat}
%\end{equation}
%Denominator of the above equation can be expressed as 
\begin{align}
&\mathbb{E}\lbrace \mathbf n_e \mathbf n_e^H\rbrace \nonumber\\
&=\mathbb{E}\lbrace [(\mathbf H^H\mathbf H)^{-1}\mathbf H^H(\mathbf n+\tfrac{1}{\alpha}\mathbf n_q)][(\mathbf H^H\mathbf H)^{-1}\mathbf H^H(\mathbf n+\tfrac{1}{\alpha}\mathbf n_q)]^H\rbrace   \nonumber\\
&=(\mathbf H^H\mathbf H)^{-1}\mathbf H^H\mathbb{E}\lbrace \underbrace{(\mathbf n+\frac{1}{\alpha}\mathbf n_q)(\mathbf n+\frac{1}{\alpha}\mathbf n_q)^H}_{\mathbf n \mathbf n^H+\frac{1}{\alpha} \mathbf n \mathbf n_q^H+\frac{1}{\alpha} \mathbf n_q \mathbf n^H+\frac{1}{\alpha^2}\mathbf n_q \mathbf n_q^H}\rbrace\mathbf H(\mathbf H^H\mathbf H)^{-1}.
\label{eq:Ene_1}
\end{align}
We assume that unquantized noise vector $\mathbf n$ and the quantization noise vector $\mathbf n_q$ are uncorrelated. %\colorbox{yellow}{validated by performing simulations.} 
Therefore, $\mathbb{E}\lbrace\mathbf n\mathbf n_q^H\rbrace$ and $\mathbb{E}\lbrace\mathbf n_q\mathbf n^H\rbrace$ are equal to zero, and (\ref{eq:Ene_1}) can be simplified as

\begin{equation}
\scriptstyle
\mathbb{E}\lbrace \mathbf n_e \mathbf n_e^H\rbrace  = (\mathbf H^H\mathbf H)^{-1}\mathbf H^H\left[\mathbb{E}\lbrace \mathbf n \mathbf n^H\rbrace+\frac{1}{\alpha^2}\mathbb{E}\lbrace\mathbf n_q \mathbf n_q^H\rbrace\right]\mathbf H(\mathbf H^H\mathbf H)^{-1},
\label{eq:Ene_2}
\end{equation}
%\begin{align}
%&\mathbb{E}\lbrace \mathbf n_e \mathbf n_e^H\rbrace\nonumber\\
%&=(\mathbf H^H\mathbf H)^{-1}\mathbf H^H\left[\mathbb{E}\lbrace \mathbf n \mathbf n^H\rbrace+\frac{1}{\alpha^2}\mathbb{E}\lbrace\mathbf n_q \mathbf n_q^H\rbrace\right]\mathbf H(\mathbf H^H\mathbf H)^{-1},
%\label{eq:Ene_2}
%\end{align}
where
\begin{align}
\mathbb{E}\lbrace\mathbf n \mathbf n^H\rbrace & = \sigma_n^2\mathbf I_N, \\
\mathbb{E}\lbrace\mathbf n_q \mathbf n_q^H\rbrace & = \alpha (1-\alpha)\diag(\sigma_x^2\mathbf H \mathbf H^H+\sigma_n^2\mathbf I_N).
\end{align}
In order to calculate the term $\diag(\cdot)$ in the above, we examine the $i$th diagonal element as
\begin{equation}
[\diag(\sigma_x^2\mathbf H \mathbf H^H +\sigma_n^2\mathbf I_N)]_{ii} =\sigma_x^2 \sum_{k=1}^K  |h_{ik}|^2 +\sigma_n^2
\end{equation}
The channel is assumed to be i.i.d Rayleigh fading, in other words, $h_{ik}$ are i.i.d complex Gaussian random variables with zero-mean and unit-variance. In \cite{lim2015performance}, it is shown that for such channel coefficients, $|h_{ik}|^2$ is a Gamma distributed random variable with unit-shape and unit-scale parameters. Equivalently, $|h_{ik}|^2$ are i.i.d exponential random variables with unit-parameter ($\lambda=1$). Furthermore, according to the \textit{weak law of large numbers} \cite{garcia2008probability}, for a fixed and large enough value of $K$, sample mean of $K$ i.i.d samples $g_k=|h_{ik}|^2$ approaches their mean value, i.e. $\lambda=1$ for our channel model. Therefore,
%We usually have $K\gg1$ for massive MIMO, and the channel is assumed to be i.i.d Rayleigh fading in this paper, therefore we can say that 
\begin{equation}
\frac{1}{K}\sum_{k=1}^K  g_k=\frac{1}{K}\sum_{k=1}^K  |h_{ik}|^2\approx 1
\end{equation}
and
\begin{equation}
\sigma_x^2 \sum_{k=1}^K  |h_{ik}|^2 +\sigma_n^2 \approx K\sigma_x^2 +\sigma_n^2.
\end{equation}

Consequently, (\ref{eq:Ene_2}) can be rewritten as
\begin{equation}
\mathbb{E}\lbrace \mathbf n_e \mathbf n_e^H\rbrace \approx
\left[\sigma_n^2+ \frac{(1-\alpha)}{\alpha}(K \sigma_x^2+\sigma_n^2)\right](\mathbf H^H\mathbf H)^{-1}
\label{eq:Ene_3}
\end{equation}
%Using (\ref{eq:Ene_3}) in (\ref{eq:SINR_xhat}), we can express the total received SINR as
%\begin{equation}
%\text{SINR} = \frac{\sigma_x^2\mathbf I_K}{\left[\sigma_n^2+ \tfrac{1-\alpha}{\alpha} (K \sigma_x^2+\sigma_n^2)\right](\mathbf H^H\mathbf H)^{-1}}
%\end{equation}
%As a result of the above equation, the received SINR of $k$th symbol stream is given by
Assuming an equal transmit power of $\sigma^2_x$ for all users, the received SINR of $k$th user is given by
\begin{align}
\gamma_{q,k}  &\approx \frac{\sigma_x^2}{\left[\sigma_n^2+ \tfrac{1-\alpha}{\alpha} (K \sigma_x^2+\sigma_n^2)\right]}\frac{1}{[(\mathbf H^H\mathbf H)^{-1}]_{kk}} \nonumber\\
              &=\gamma_{q0}\frac{1}{[(\mathbf H^H\mathbf H)^{-1}]_{kk}}
\label{eq:gamma_k_q}
\end{align}
As explained in the previous section, $\chi^2_d={1}/{[(\mathbf H^H\mathbf H)^{-1}]_{kk}}$ is a chi-square distributed random variable with $d=2(N-K+1)$ degrees of freedom. Furthermore, we assume the SINR per streams for all users are identically distributed. Therefore, the SINR of each symbol stream is represented by $\gamma_q=\gamma_{q0} \chi^2_d$, where
\begin{equation}
\gamma_{q0}= \frac{\sigma_x^2}{\left[\sigma_n^2+ \tfrac{1-\alpha}{\alpha} (K \sigma_x^2+\sigma_n^2)\right]}.
\label{eq:gamma_q0}
\end{equation}

%------------------------------------------------------ Quantized Massive MIMO
\section{BER of Quantized M-QAM Massive MIMO}
\label{sec:BER_QuantizedMassiveMIMO}

\subsection{BER of M-QAM MIMO}
An analytical BER expression of an i.i.d. Rayleigh fading MIMO, for square M-QAM modulations (with Gray code mapping and ZF detection) is obtained in \cite{wang2007performance} by averaging over the bit error probability with respect to the chi-squared random variable $\chi^2_d$ that is addressed in the previous subsection. Readily, the BER of an M-QAM MIMO base station is given by \cite{wang2007performance}
\begin{align}
&BER_{MQAM}  \cong \frac{2}{\sqrt{M}\log_2\sqrt{M}}\sum_{k=1}^{\log_2\sqrt{M}} \hspace{0.5em}\sum_{i=0}^{(1-2^{-k})\sqrt{M}-1}
\nonumber\\
&\Biggl\lbrace (-1)^{\lfloor\frac{i.2^{k-1}}{\sqrt{M}}\rfloor} \left(\ 2^{k-1}- \lfloor \frac{i.2^{k-1}}{\sqrt{M}}+\frac{1}{2}\rfloor\right) B(i)\Biggr\rbrace
\label{eq:BER_MQAM}
\end{align}
where $\lfloor x\rfloor$ is the floor function giving the greatest integer, less than or equal to the input $x$, and
\begin{equation}
B(i)={[\frac{1}{2}(1-\mu_i)]}^{D+1}\sum_{j=0}^{D} \binom {D+j}j {[\frac{1}{2}(1+\mu_i)]}^j
\label{eq:Bi}
\end{equation}
where
\begin{equation}
\mu_i  = \sqrt{\frac{3(2i+1)^2\gamma_{0} }{2(M-1)+3(2i+1)^2\gamma_{0}}},\hspace{1em}  D=N-K.
\label{eq:mu_i}
\end{equation}

The BER estimate of M-QAM in (\ref{eq:BER_MQAM}), can be more simplified by keeping only two dominant terms at $i=0,1$ and neglecting the rest. Therefore we have \cite{wang2007performance} 
\begin{equation}
BER_{MQAM} \cong \frac{2(\sqrt{M}-1)}{\sqrt{M}\log_2\sqrt{M}}B(0)+\frac{2(\sqrt{M}-2)}{\sqrt{M}\log_2\sqrt{M}}B(1)
\label{eq:BER_MQAM_simple}
\end{equation}
The above analytical BER formula is validated in \cite{wang2007performance} for M-QAM MIMO, and here we use it for the case of unquantized massive MIMO.

%\subsection{Quantization Resolution Effects on BER Performance of M-QAM Massive MIMO}
\subsection{BER of Quantized M-QAM Massive MIMO}
\label{sec:BER_q}
A closed-form BER expression of a massive MIMO base station using low-resolution ADCs, for M-QAM modulations and ZF detection, can be obtained from (\ref{eq:BER_MQAM_simple}) if we replace $\gamma_{0}$ by $\gamma_{q0}$ from (\ref{eq:gamma_q0}).

%\section{Asymptotic BER Performance of Quantized M-QAM Massive MIMO}
%\label{sec:AsymptoticBER}
We further investigate the effects of quantization on the BER performance of uplink massive MIMO, by considering the following cases.

\subsection{Increasing the bit resolution of ADCs}
As $b$ goes to infinity, $\alpha$ approaches 1. Then,
\begin{equation}
\gamma_{q0}=\frac{\sigma_x^2}{\left[\sigma_n^2+ \tfrac{1-\alpha}{\alpha} (K \sigma_x^2+\sigma_n^2)\right]}
\xrightarrow[\quad b\to\infty \quad]{} \gamma_0=\frac{\sigma_x^2}{\sigma_n^2}.
\end{equation}	
Accordingly, $\gamma_{q,k}$ in (\ref{eq:gamma_k_q}) approaches $\gamma_k$ in (\ref{eq:gamma_k}), and we will have the same received SINR for both quantized and unquantized massive MIMO.

\subsection{Increasing the transmit power}
As we increase the transmit power to infinity, for an unquantized massive MIMO system, $\gamma_0$ approaches $\infty$ and therefore the BER performance goes to zero. However, for the quantized massive MIMO, we have 
\begin{equation}
\gamma_{q0}\xrightarrow[\quad \left({\sigma_x^2}/{\sigma^2_n}\right)\to\infty \quad]{} \frac{\alpha}{(1-\alpha)K},
\end{equation}
and BER goes to a non-zero constant value. We readily see that this BER floor can be reduced if we increase $\gamma_{q0}$ either by increasing the quantization resolution that result in $\alpha\to 1$, or reducing the number of users K.
	
\subsection{Increasing the number of users, K}
Referring to (\ref{eq:gamma_k_q}), as we increase the number of users, the denominator of $\gamma_{q,k}$ (representing the sum of quantization noise and interferences) is increased. Therefore, the BER gets worse by increasing $K$ for quantized massive MIMO.
%Furthermore, the degrees of freedom of $\chi^2_d$ is decreased by increasing $K$.

%------------------------------------------------------ Numerical Result
\section{Numerical Results}
\label{sec:Simulations}
In this section, some numerical simulations are performed to investigate the accuracy of the proposed analytical BER expression for coarse quantized massive MIMO systems. 
We consider an uplink massive MIMO with $N=100$ antennas at the BS and $K=10$ users, employing two types of \emph{uniform} and \emph{non-uniform} quantizers, with different quantization bit resolutions of $b=1,2,3,4$ and full precision (i.e., $b=\infty$).

\subsection{Uniform Quantization}
\begin{figure}[t]
	\centering
	\begin{subfigure}{0.99\linewidth}
		\includegraphics[width=\linewidth]{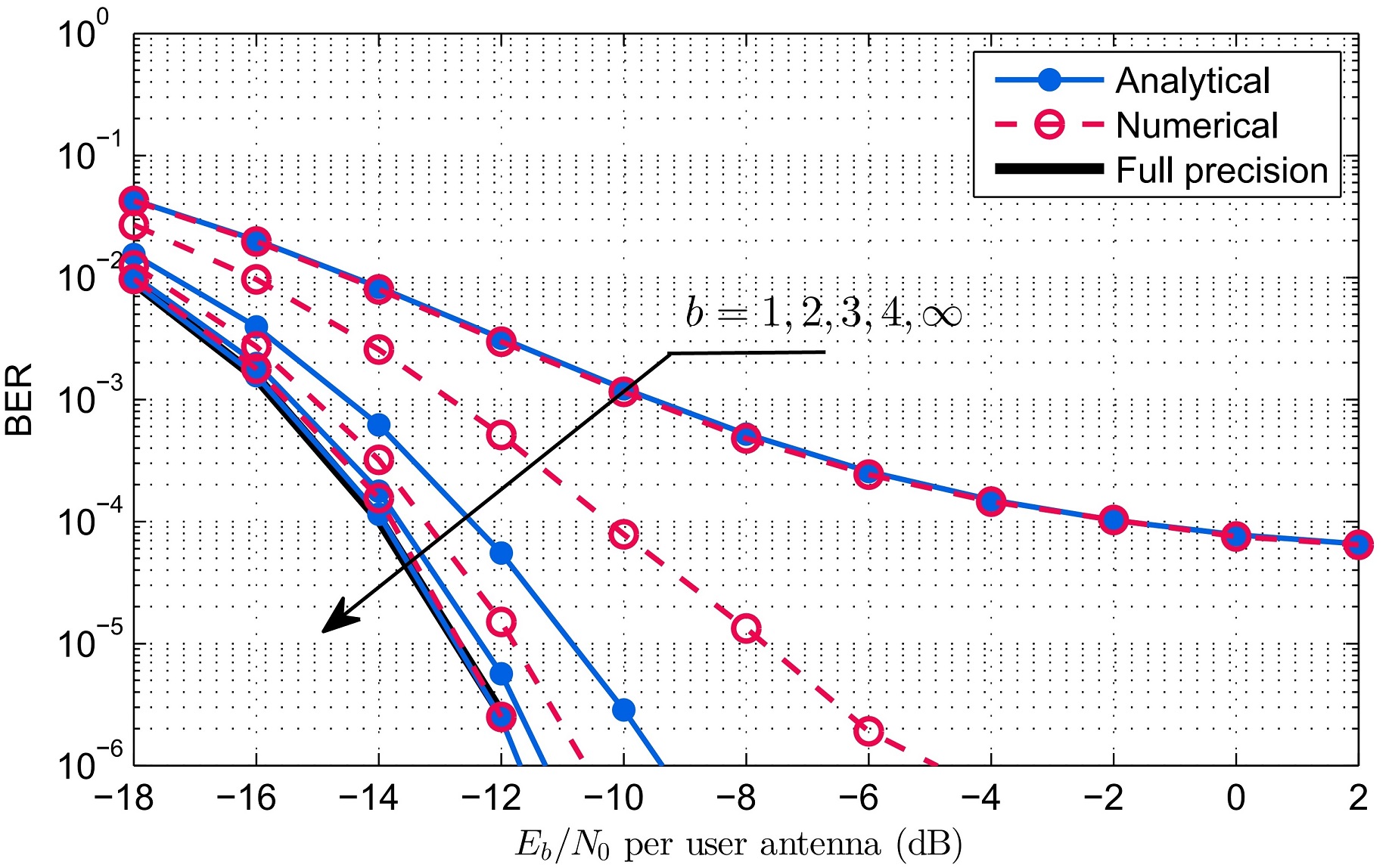}
		\caption{QPSK}
		\label{fig:BER_QPSK_uniform}
	\end{subfigure}
	\vspace{1em}
	\begin{subfigure}{0.99\linewidth}
		\includegraphics[width=\linewidth]{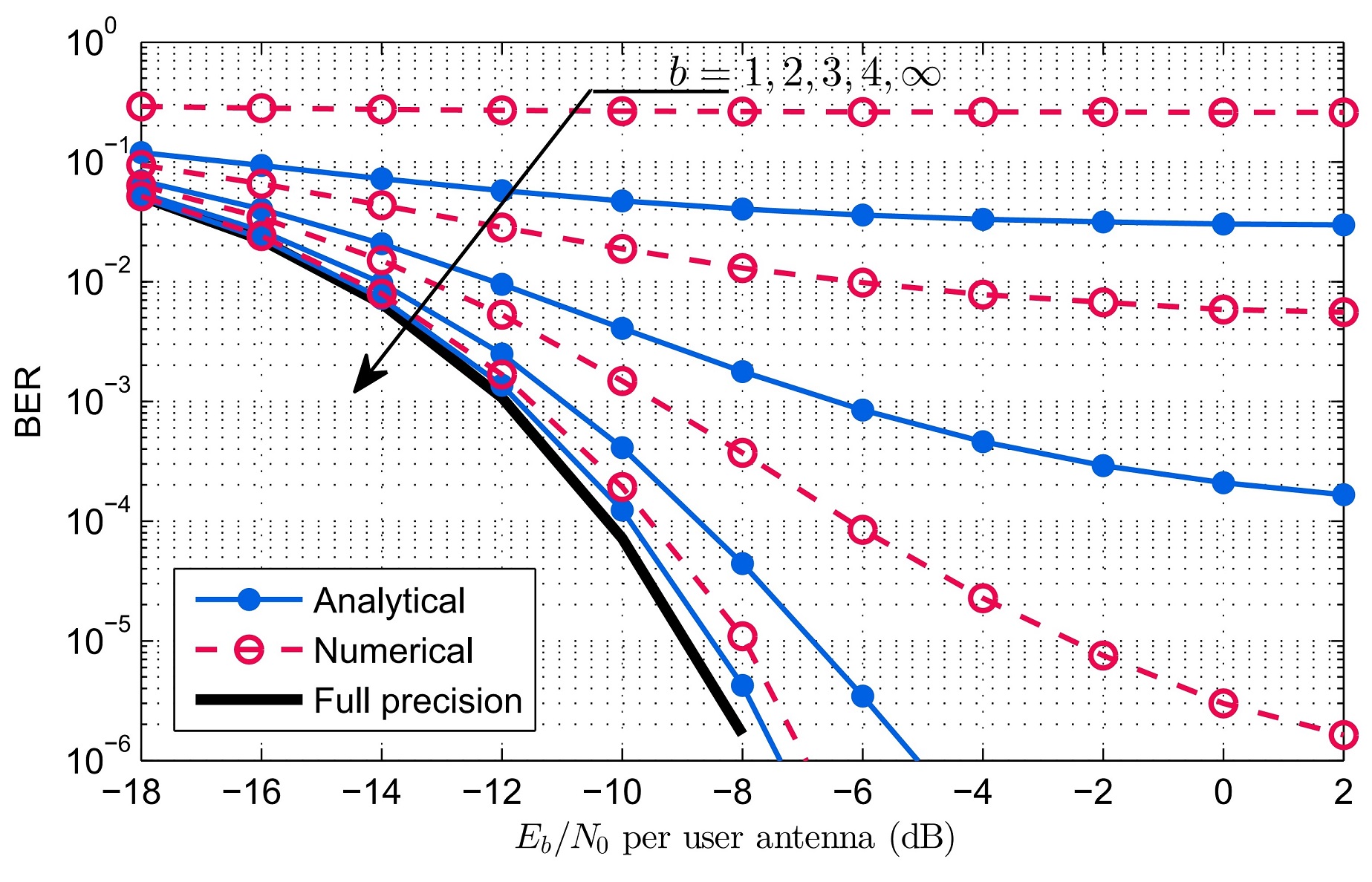}
		\caption{16-QAM}
		\label{fig:BER_16QAM_uniform}
	\end{subfigure}
	\caption{BER of quantized massive MIMO for (a) QPSK, and (b) 16-QAM modulations with $N=100$, and $K=10$, using uniform quantizer.}
	\label{fig:BER_QPSK_16QAM_uniform}
\end{figure}

We numerically analyze the BER performance of quantized QPSK and 16-QAM modulations, employing uniform quantizers, in Fig. \ref{fig:BER_QPSK_16QAM_uniform}. Numerical results are compared to the corresponding BER values obtained from the \emph{analytical} formula of (\ref{eq:BER_MQAM_simple}) that uses the \emph{linear quantization model}.  

Looking at the BER curves, we observe that analytical curves give an upper bound for the BER performance of the corresponding numerical curves, with noticeable gaps between numerical and analytical curves for both QPSK and 16-QAM modulations at very low-resolution quantization ($b=1,2,3$). With an exception of 1-bit QPSK that both curves are matched, discrepancies are arising by increasing $E_b/N_0$ (SNR per bit) per user. 
This may happen due to the inaccuracy of the linear quantization model for uniform quantizers at low-bit resolutions \cite{fletcher2007robust}. Therefore, we examine the use of non-uniform quantizers at the following subsection.

%Another point is that numerical BER curves are worse in comparison with the corresponding analytical curves. 
%However, there is an exception that both numerical and analytical curves are matched very well for 1-bit quantized QPSK???????????????

%\begin{figure}[t]
%	\centering
%	\includegraphics[width=0.99\linewidth]{Figures/fig3.jpg}
%	\caption{Decision boundaries of a standard normal distributed input signal $y$ and output points $y_q$ for a 2-bits non-uniform quantizer.}
%	\label{fig:non-uniform_quantizer}
%\end{figure}

\subsection{Non-uniform Quantization}
Since the linear quantization model parameters in Table \ref{tab:rho} is provided for a non-uniform quantizer \cite{max1960quantizing}, we regenerate our numerical BER curves by using a non-linear quantizer described in \cite{max1960quantizing} that is optimal for a Gaussian distributed input signal. 

From now on, we use the above \emph{non-uniform quantizer} for the rest of numerical results. 
%As an example, decision intervals of a standard normal input signal $y$ and the corresponding output points $y_q$ for a 2-bit quantizer are illustrated in Fig. \ref{fig:non-uniform_quantizer}.

\begin{figure}[t]
	\centering
	\begin{subfigure}{0.99\linewidth}
		\centering
		\includegraphics[width=\linewidth]{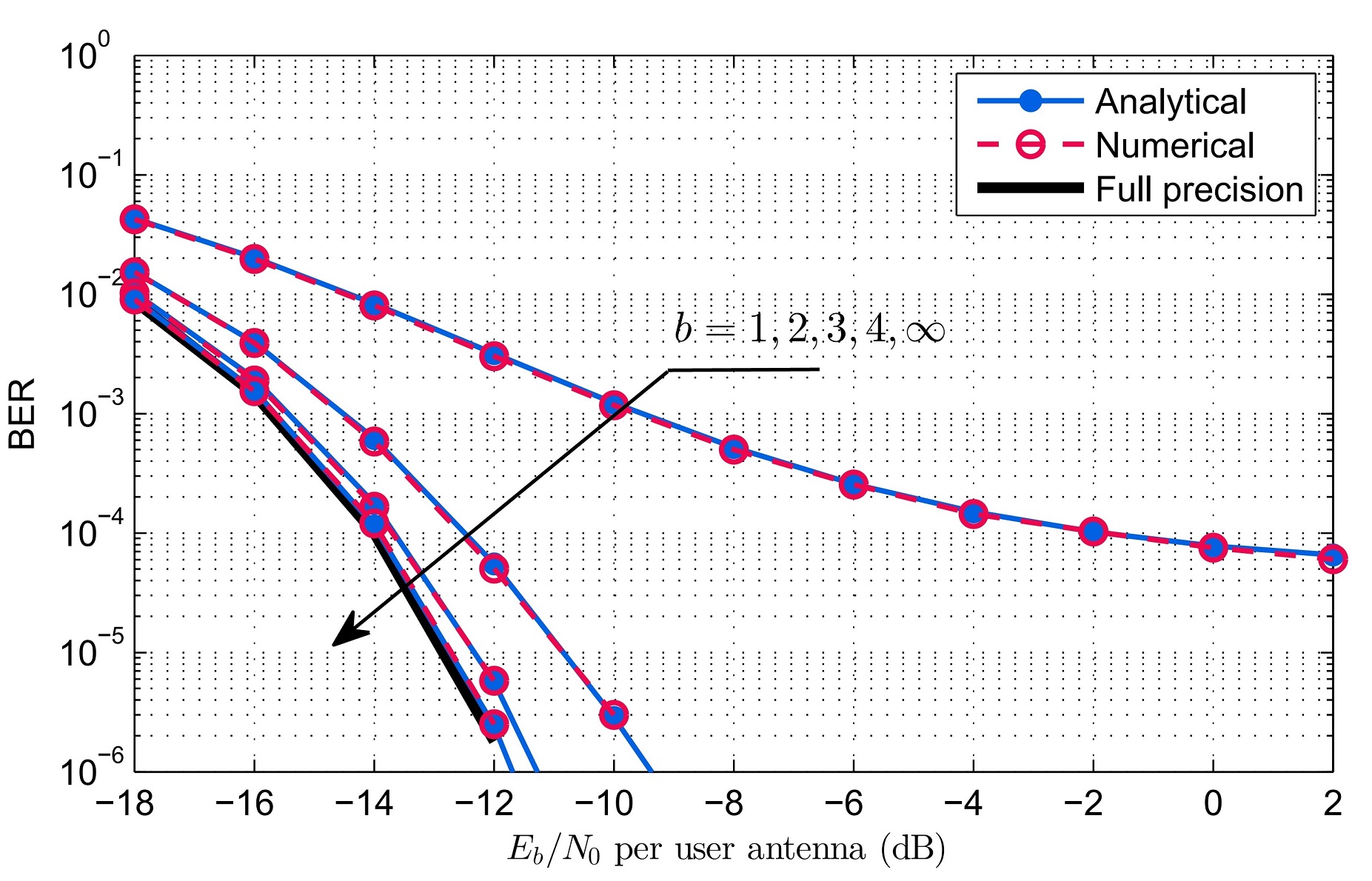}
		\caption{QPSK}
		\label{fig:BER_QPSK_Non-uniform}
	\end{subfigure}
~
	\begin{subfigure}{0.99\linewidth}
		\centering
		\includegraphics[width=\linewidth]{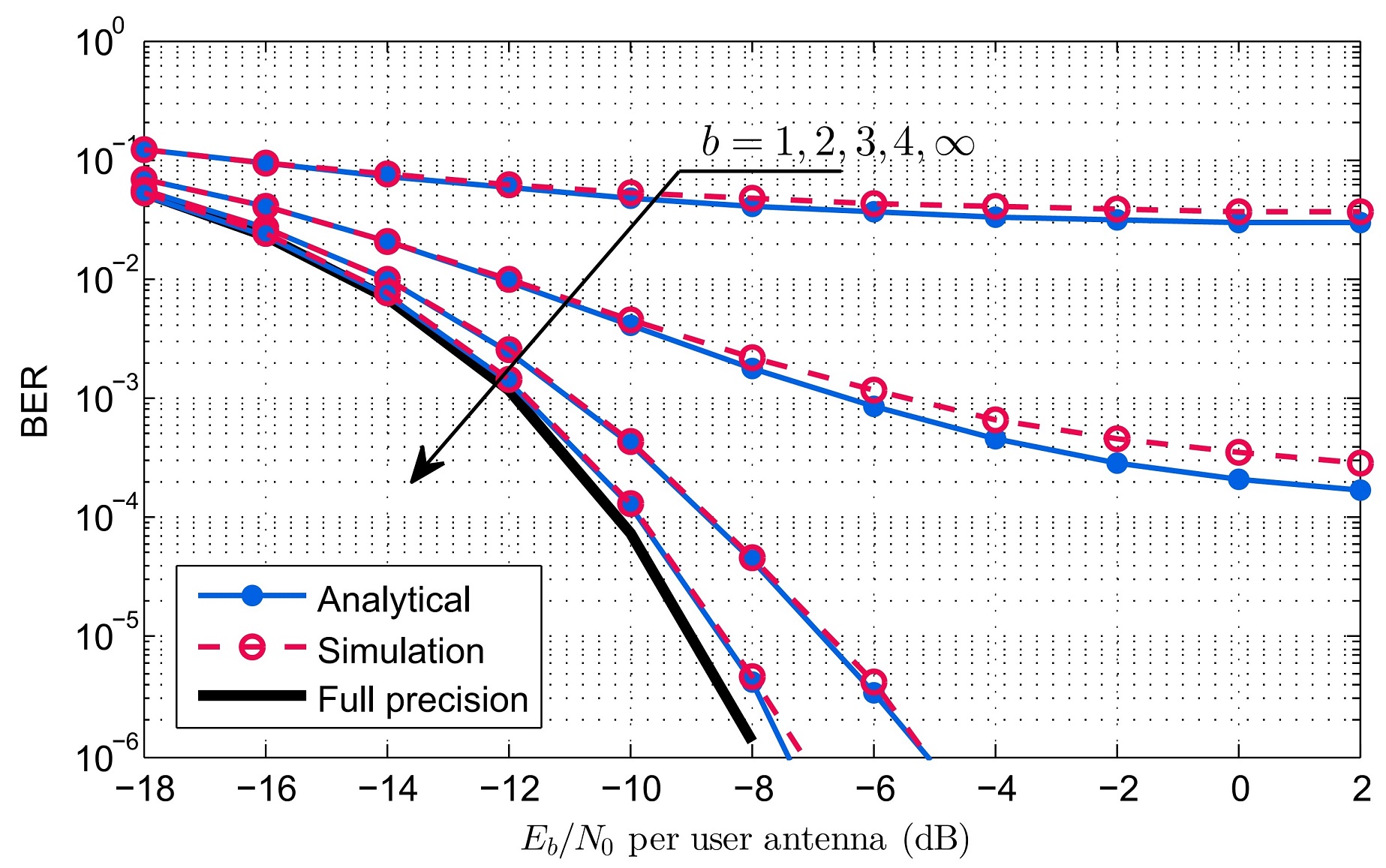}
		\caption{16-QAM}
		\label{fig:BER_16QAM_Non-uniform}
	\end{subfigure}
~
%\end{figure}
%\begin{figure}[t]\ContinuedFloat
%
\begin{subfigure}{0.99\linewidth}
	\centering
	\includegraphics[width=\linewidth]{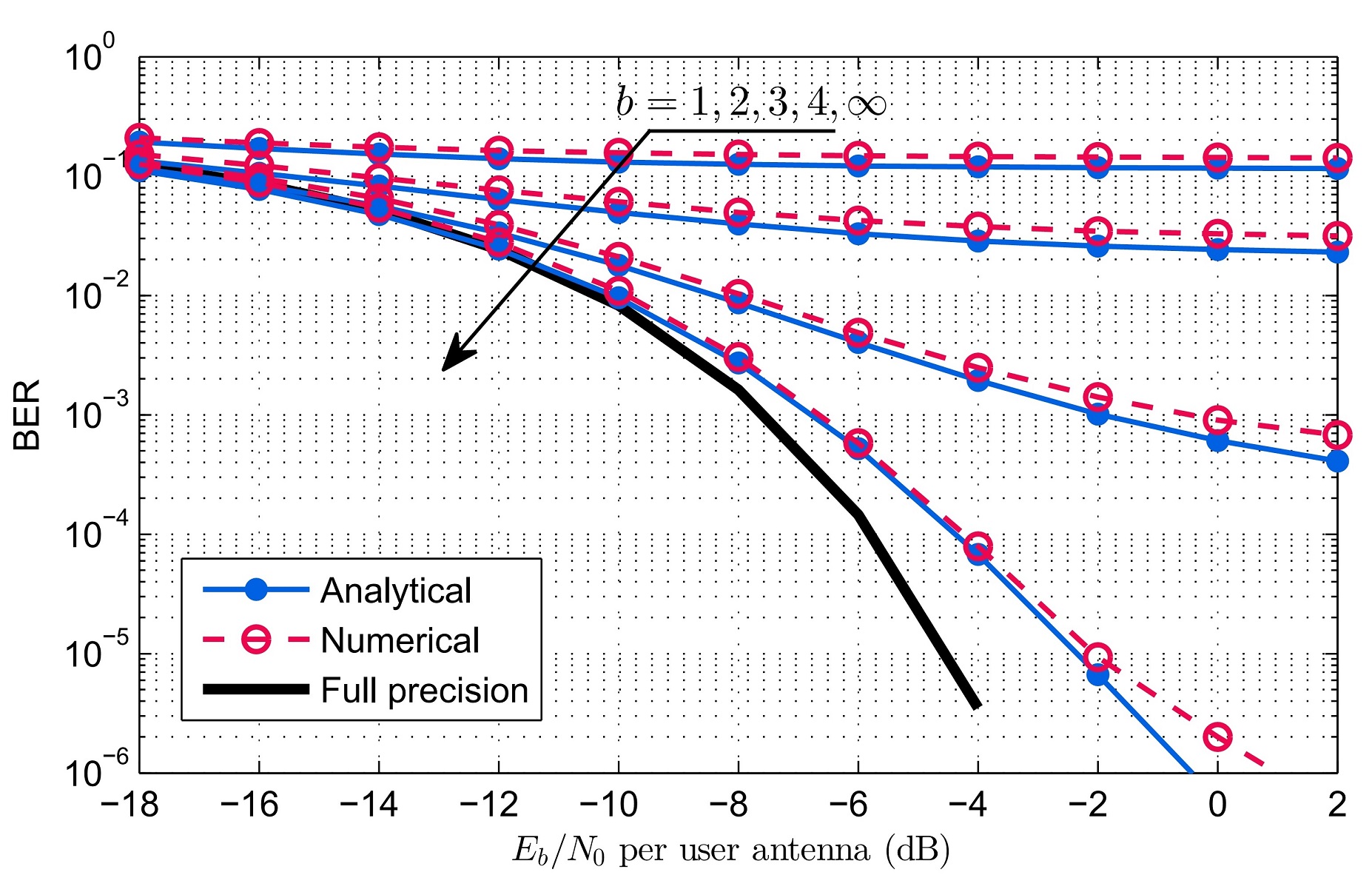}
	\caption{64-QAM}
	\label{fig:BER_64QAM_Non-uniform}
\end{subfigure}
	\caption{BER of quantized massive MIMO for (a) QPSK, (b) 16-QAM, and (c) 64-QAM modulations with $N=100$, and $K=10$, using non-uniform quantizer.}
	\label{fig:BER_MQAMs_Non-uniform}
\end{figure}

The numerical BER performance of three modulation types of QPSK, 16-QAM and 64-QAM are illustrated in Fig. \ref{fig:BER_MQAMs_Non-uniform}.  As we see, the numerical and analytical BER curves for QPSK are similar, even for very low quantization resolutions of $b=1$, and 2-bits. Furthermore, we observe a very small difference between numerical and analytical values of BER at $b=1,2$ for 16-QAM, and $b=1,2,3$ for 64-QAM. However, the gap between these curves is slowly growing by increasing the SNR per bit ($E_b/N_0$).

Therefore, we see that analytical BER expression provides a very tight upper bound for the BER performance of coarse quantized systems having non-uniform quantizers.

Another lessen learned from Fig. \ref{fig:BER_MQAMs_Non-uniform} is that a very poor BER performance is achieved for the coarse quantized cases of (i) 1-bit 16-QAM, (ii) 1-bit and 2-bits 64-QAM. Therefore, we need to increase the bit-resolution of quantization. However, the effects of coarse quantization might be somehow compensated by employing coding techniques.

\begin{table*}[h]
	\centering
	\resizebox{0.98\textwidth}{!}{\begin{minipage}{\textwidth}
			%\begin{table*}
			\renewcommand{\arraystretch}{1.3}
			\centering
			\caption{BER values  for SNR$\to\infty$  in an uncoded M-QAM massive MIMO system with $K=10$ users and $N=100$ antennas at the BS, having $b$-bit resolution ADCs.}
			\label{tab:BER_snr_inf}
			\begin{tabular}{|c||c|c||c|c||c|c||c|c||}\hline
				\multirow{2}{*}{}& \multicolumn{2}{|c||}{$b=1$}& 	\multicolumn{2}{|c||}{$b=2$}& \multicolumn{2}{|c||}{$b=3$}&\multicolumn{2}{|c||}{$b=4$}\\ \hline
				 &Analytical&Numerical                &	Analytical&Numerical&	Analytical&Numerical&	Analytical&Numerical\\ \hline\hline 			
				QPSK& $4.76\times10^{-5}$&$4.8\times10^{-5}$&	$1.4\times10^{-14}$&0&	$1.1\times10^{-36}$&0&	$2.47\times10^{-74}$&0\\ \hline	
				16-QAM&	$2.84\times10^{-2}$&$3.52\times10^{-2}$&	$1.8\times10^{-4}$&$2.31\times10^{-4}$&	$8.5\times10^{-12}$&0&	$1.91\times10^{-30}$&0\\ \hline
				64-QAM&	$1.14\times10^{-1}$&$1.41\times10^{-1}$&	$2.13\times10^{-2}$&$2.85\times10^{-2}$&	$1.83\times10^{-4}$&$3.69\times10^{-4}$&	$6.59\times10^{-11}$&$2.33\times10^{-9}$\\ \hline
			\end{tabular}
	\end{minipage} }
\end{table*}

%\begin{table*}[h]
%	\centering
%	\resizebox{0.95\textwidth}{!}{\begin{minipage}{\textwidth}
%			%\begin{table*}
%			\renewcommand{\arraystretch}{1.3}
%			\centering
%			\caption{BER values  for SNR$=0$ dB in an uncoded M-QAM massive MIMO system with $K=10$ users and $N=100$ antennas at the BS, having $b$-bit resolution ADCs.}
%			\label{tab:BER_snr_zero}
%			\begin{tabular}{|c||c|c||c|c||c|c||c|c||}\hline
%				& \multicolumn{2}{|c||}{$b=1$}& 	\multicolumn{2}{|c||}{$b=2$}& \multicolumn{2}{|c||}{$b=3$}&\multicolumn{2}{|c||}{$b=4$}\\ \hline
%				&	Analytical&Numerical                &	Analytical&Numerical&	Analytical&Numerical&	Analytical&Numerical\\ \hline\hline 			
%				QPSK& $7.41\times10^{-5}$ & $7.8\times10^{-5}$&	$1.3\times10^{-12}$ &0 & $1.5\times10^{-25}$&0 &	$2.52\times10^{-37}$&0\\ \hline	
%				16-QAM&	$3.03\times10^{-2}$&$3.66\times10^{-2}$& $2.08\times10^{-4}$&$3.75\times10^{-4}$&	$2.19\times10^{-9}$&0&	$7.97\times10^{-17}$&0\\ \hline
%				64-QAM&	$1.16\times10^{-1}$&$1.43\times10^{-1}$&	$2.42\times10^{-2}$&$3.17\times10^{-2}$&	$6.06\times10^{-4}$& $8.9\times10^{-4}$&	$5.98\times10^{-7}$&$1.9\times10^{-6}$\\ \hline
%			\end{tabular}
%	\end{minipage} }
%\end{table*}

\begin{figure}[t]
	\centering
	\includegraphics[width=0.98\linewidth]{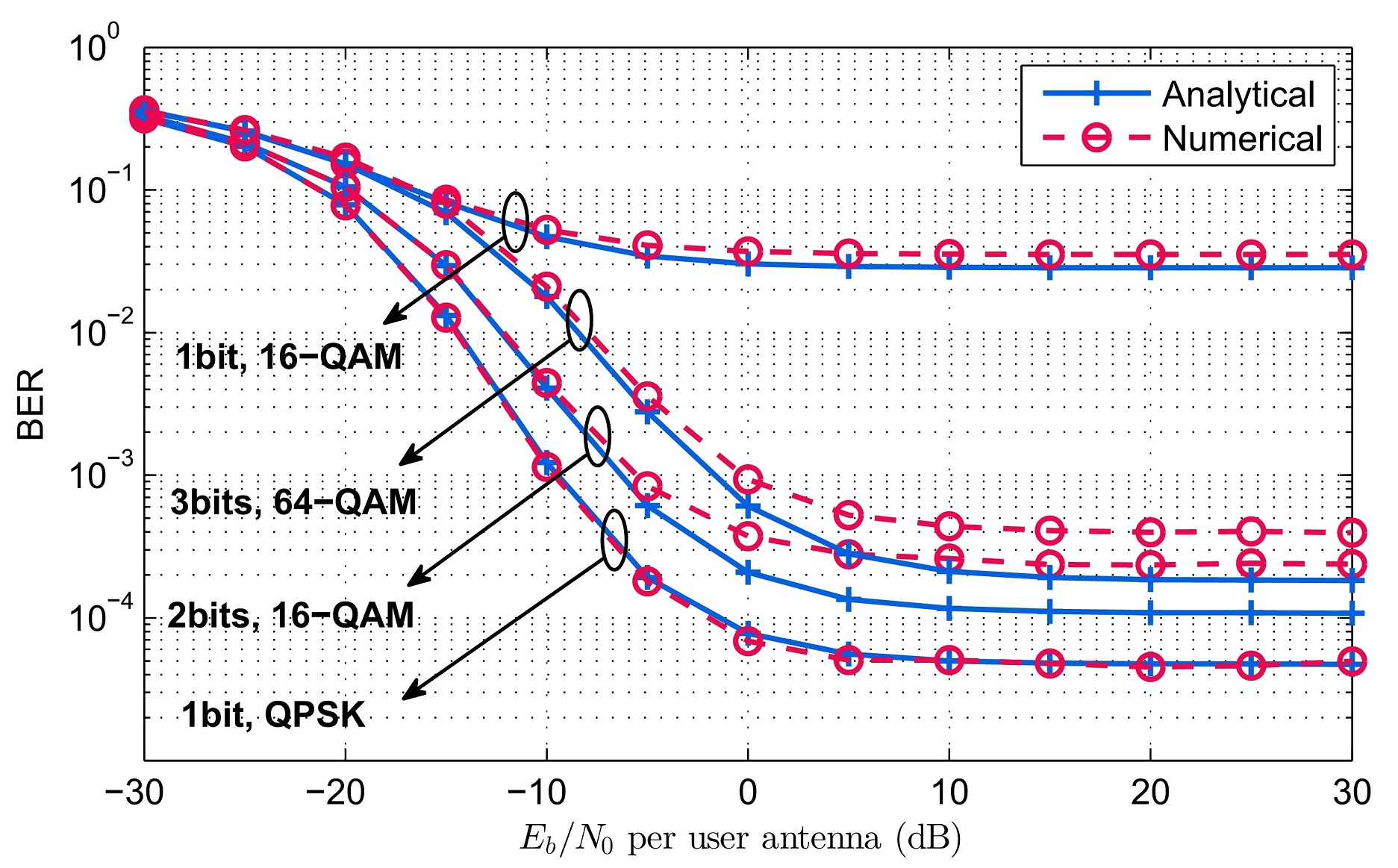}
	\caption{BER of coarse quantized massive MIMO with $N=100$, $K=10$ and modulation types of QPSK, 16-QAM, and 64-QAM.}
	\label{fig:BER_floor_MQAMs}
\end{figure}

\begin{figure}[t]
	\centering
	\includegraphics[width=0.98\linewidth]{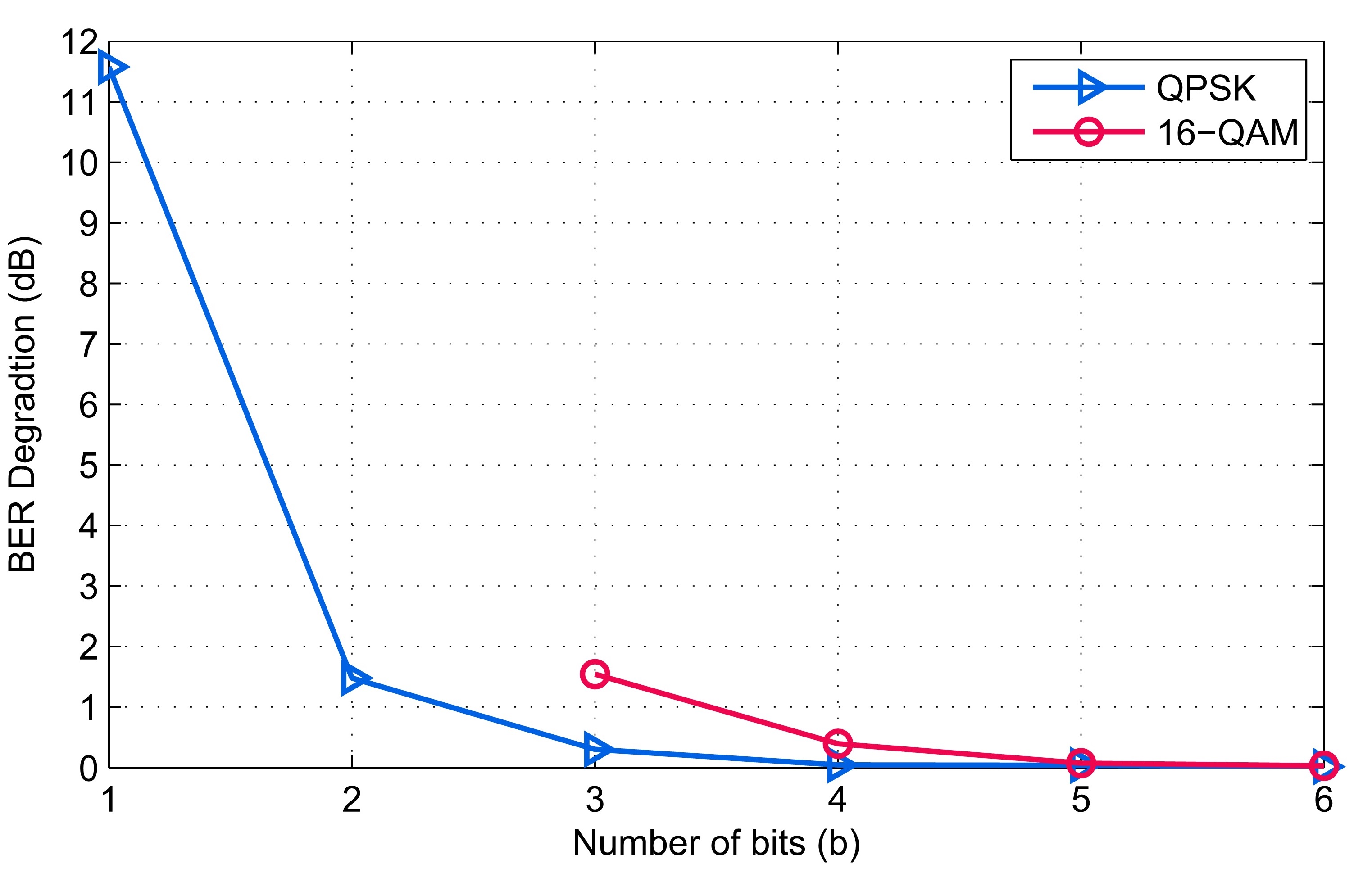}
	\caption{BER degradation as a function of $b$-bit quantization resolution for a massive MIMO system with $N=100$, and $K=10$, using QPSK and 16-QAM modulations.}
	\label{fig:BER_degradation_b}
\end{figure}

\subsection{BER Floor at Low-Resolution Quantization}
As we discussed earlier in Section \ref{sec:BER_QuantizedMassiveMIMO}, the BER of unquantized system approaches zero by increasing the SNR. However, the BER of a low-resolution quantized system goes to a non-zero value and we can not further decrease it by increasing the transmit power. 

As we observed in Fig. \ref{fig:BER_MQAMs_Non-uniform}, we can improve the BER performance and achieve very low BER values by having a very small increase in bit-resolution of quantizers, from 1-bit to 3-bits even in 64-QAM. 
Therefore, Fig. \ref{fig:BER_floor_MQAMs} demonstrates the following coarse quantized modulations: (i) QPSK at $b=1$, (ii) 16-QAM at $b=1,2$, and (iii) 64-QAM at $b=3$-bits quantization resolution. 
We see that, in the case of 1-bit 16-QAM, we can not reach a BER of $10^{-2}$ or lower. 
In addition, a similar trend is observed for both numerical and analytical BER curves of 1-bit QPSK, 2-bits 16-QAM, and 3-bits 64-QAM with a BER floor of $10^{-4}$, with QPSK and 64-QAM attaining the lowest and the worst BER, respectively. It means that, higher bit-resolution is needed to achieve the same BER performance by using higher order modulations.  
Moreover, the gap between the numerical (using non-uniform quantizer) and the analytical curves of the BER performance, for the two coarse quantized modulations of 16-QAM and 64-QAM, is growing by increasing the SNR. This issue is also reported in \cite{orhan2015low} for the capacity of low-resolution quantized massive MIMO. This might happen due to the inaccuracy of  the linear quantizer model for high SNR values.

Table \ref{tab:BER_snr_inf} provides BER values of $b$-bit quantized systems employing uncoded modulations of QPSK, 16-QAM, and 64-QAM when the \emph{SNR per user} goes to infinity. Analytical results are obtained by using asymptotic formulas in Section \ref{sec:BER_QuantizedMassiveMIMO}, and numerical results are coming from Monte Carlo simulations having a very high SNR of 100 dB.
As we see from the table, analytical BER values show an upper bound (best possible value) of the BER performance for any  M-QAM modulation, using $b$-bit resolution quantizers. This will help to decide easily what number of bit-resolution would be required.

\begin{figure}[t]
	\centering
	\begin{subfigure}{0.98\linewidth}
		\includegraphics[width=\linewidth]{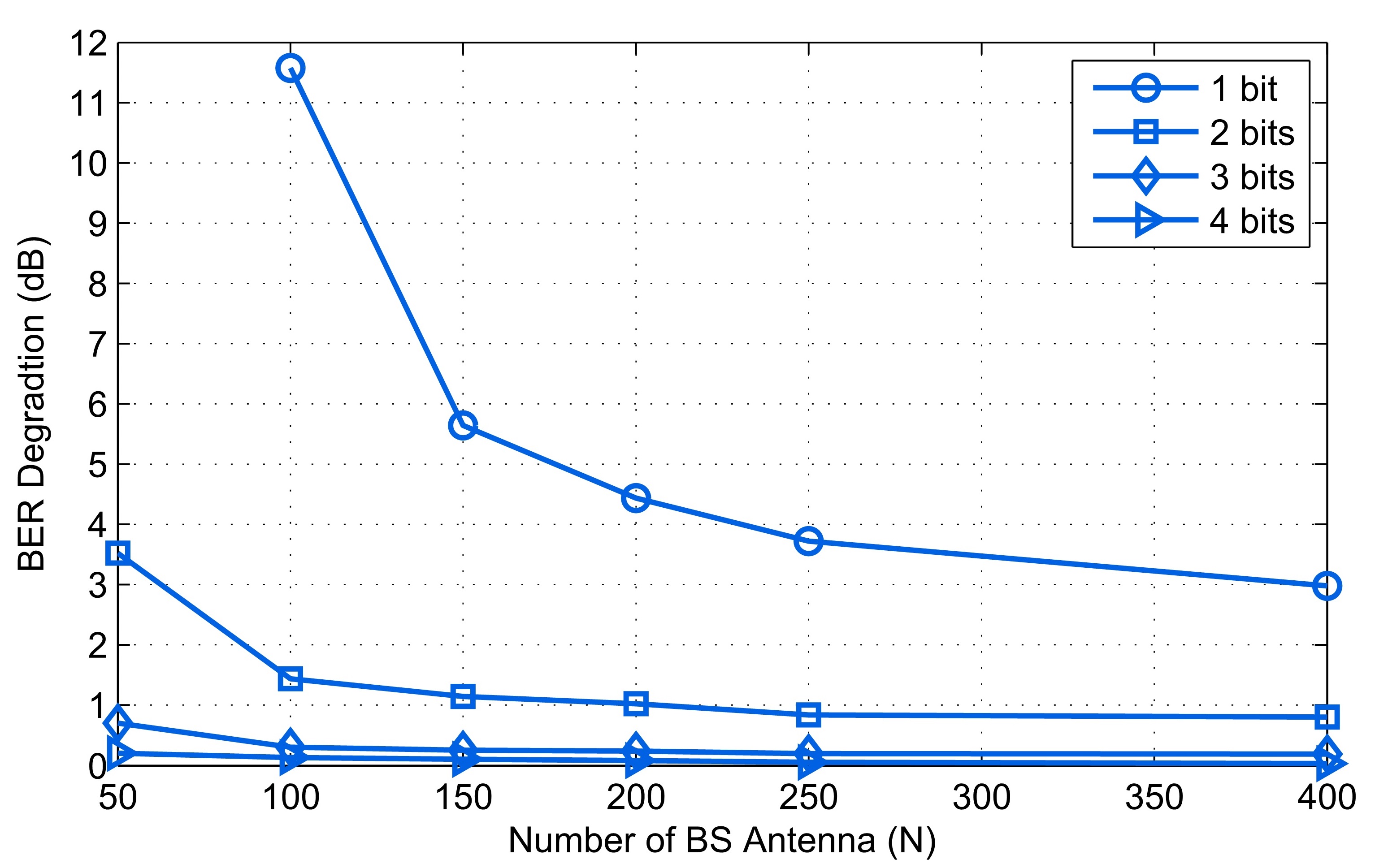}
		\caption{QPSK}
		\label{fig:BER_degrad_N_QPSK}
	\end{subfigure}
	\vspace{1em}
	\begin{subfigure}{0.98\linewidth}
		\includegraphics[width=\linewidth]{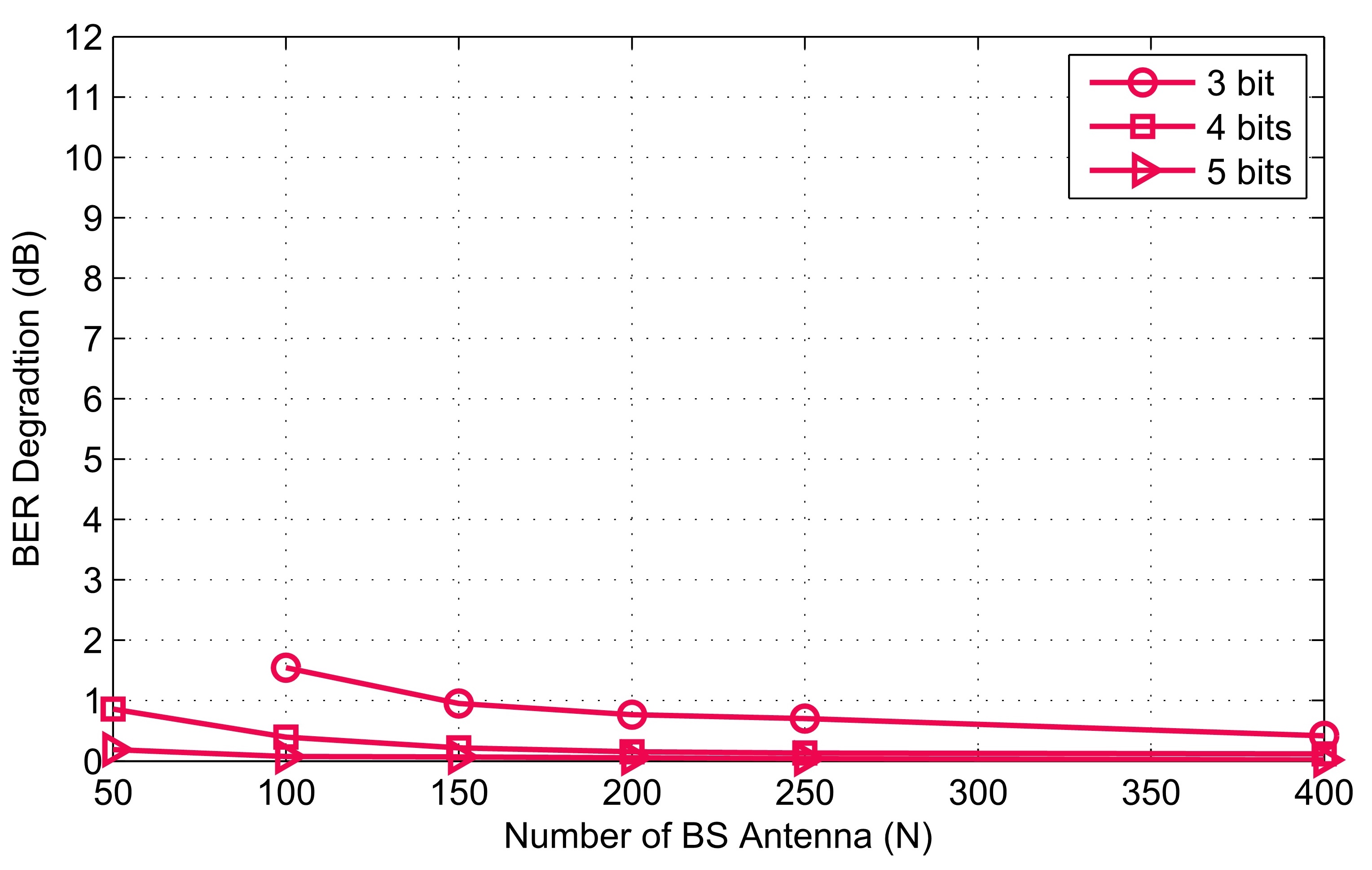}
		\caption{16-QAM}
		\label{fig:BER_degrad_N_16QAM}
	\end{subfigure}
	\caption{BER degradation of QPSK and 16-QAM modulations with $K=10$, and different number of BS antennas ($N$).}
	\label{fig:BER_degrad_N_QPSK-16QAM}
\end{figure}

\subsection{BER Degradation}
Considering a reference SNR for the unquantized (full-precision) system to achieve a BER of $10^{-4}$, we calculate the extra SNR (in dB) required to attain the same BER for $b$-bit resolution quantized system, and we call it \emph{BER degradation}. We note that this process is performed separately for each types of modulations.

Fig. \ref{fig:BER_degradation_b} shows the BER degradation of QPSK and 16-QAM modulations in quantized massive MIMO systems with $b$-bit ADCs' resolution. We see that 2-bits QPSK and 3-bits 16-QAM have the same BER degradation value of roughly 1.5 dB, whereas this value is reported in \cite{azizzadeh2017ber} to be 3 to 4 dB larger for uniform quantization. Furthermore, 3-bits QPSK and 4-bits 16-QAM have a very little BER degradation of 0.3 dB. Comparing these results to the corresponding values for uniform quantization in \cite{azizzadeh2017ber}, we see that optimal non-uniform quantizers achieve much higher BER performance at coarse quantized systems, although we will have similar performance for both non-uniform and uniform quantizations at higher resolutions.

We further investigate the effects of quantization bit-resolution and the number of BS antennas, on the BER degradation in Fig. \ref{fig:BER_degrad_N_QPSK-16QAM} for QPSK and 16-QAM modulations. It can be seen that, the BER degradation caused by lower quantization resolution, can be compensated by increasing the number of antennas at the BS for both QPSK and 16-QAM modulations.

\begin{figure}[t]
	\centering
%	\resizebox{0.5\textwidth}{!}{\begin{minipage}{\textwidth}
	\begin{subfigure}{0.85\linewidth}
		\centering
		\includegraphics[width=\linewidth]{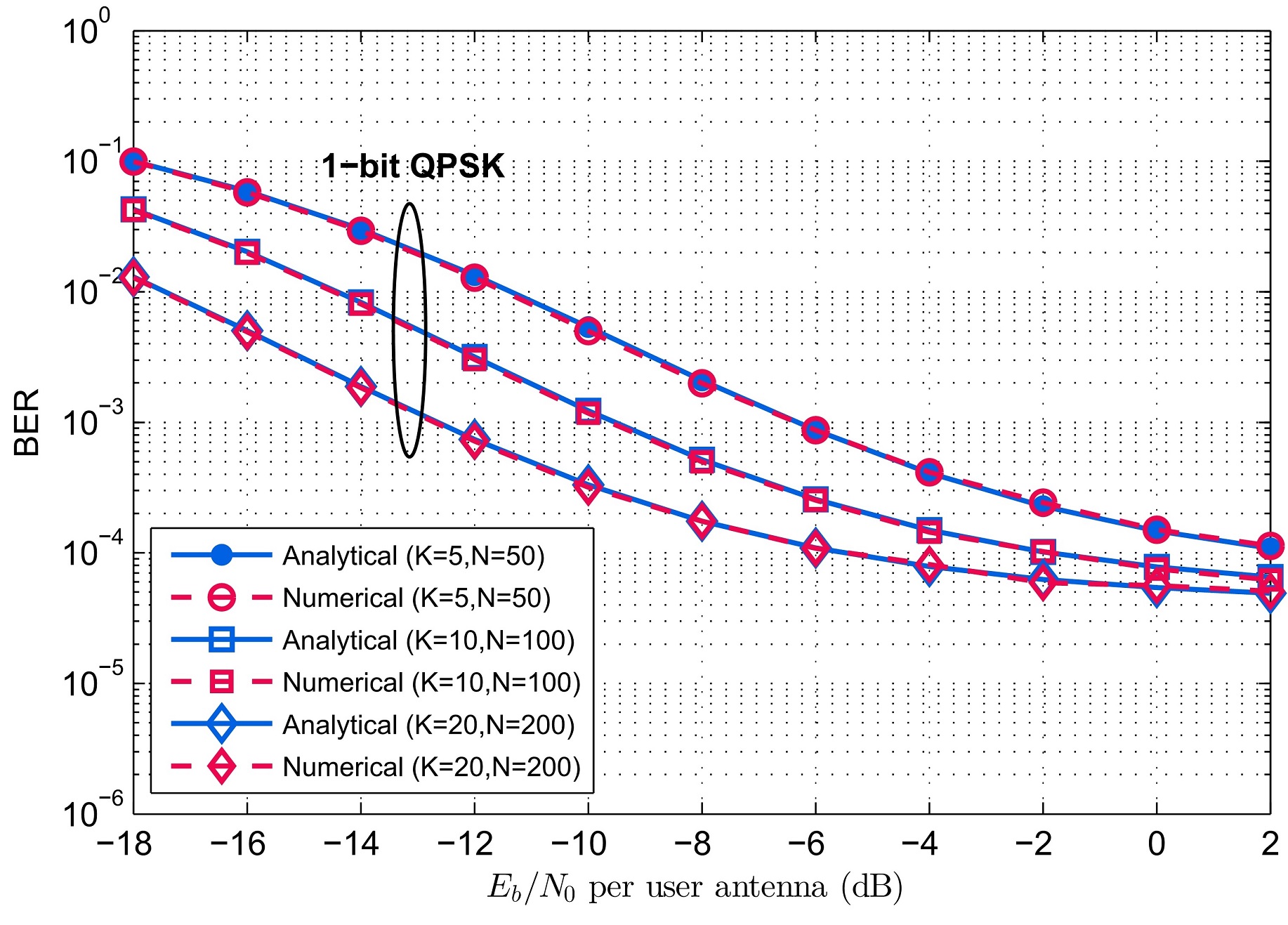}
		\caption{QPSK}
		\label{fig:BER_K_N_b1_QPSK}
	\end{subfigure}
	\begin{subfigure}{0.85\linewidth}
		\centering
		\includegraphics[width=\linewidth]{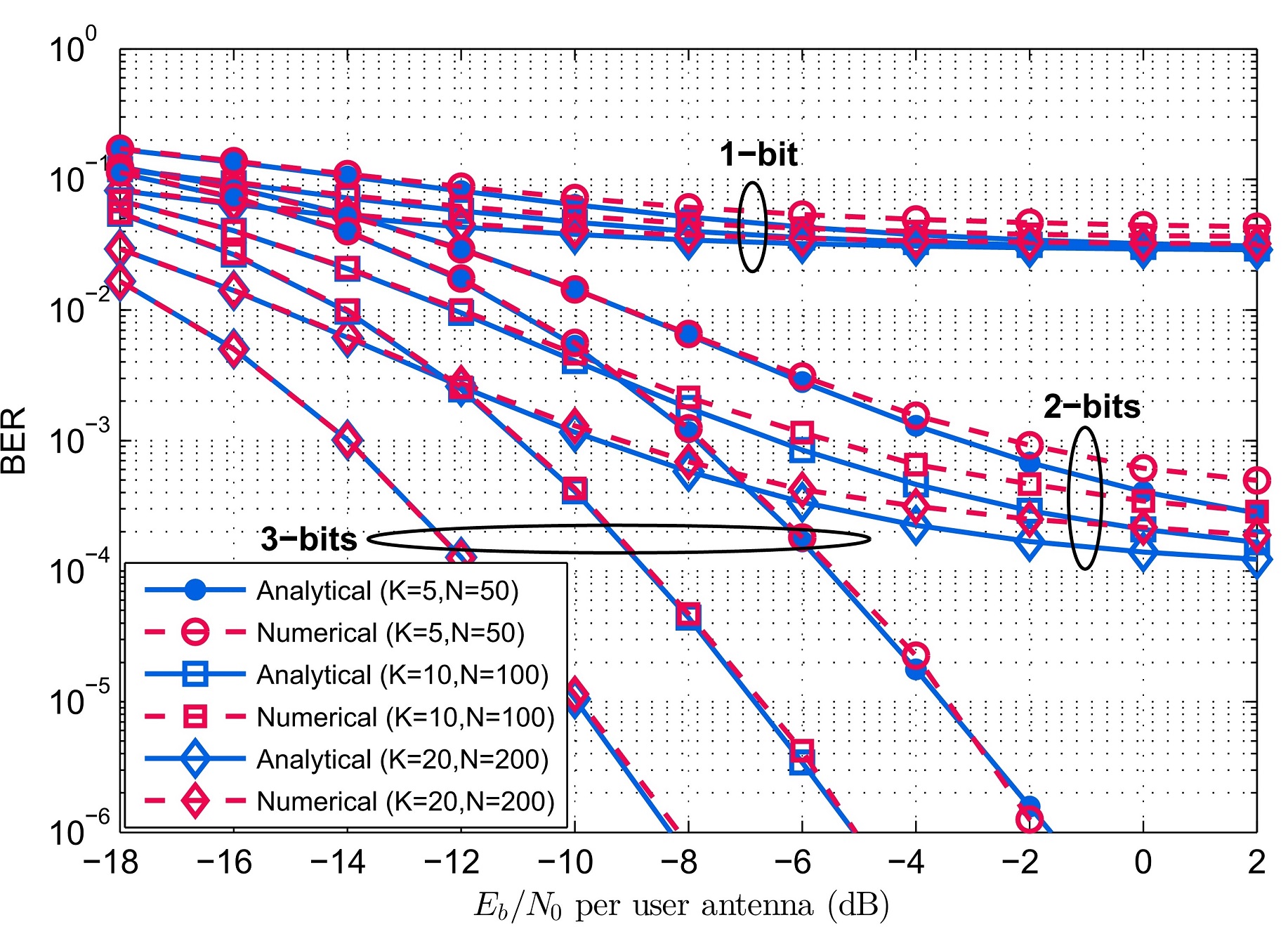}
		\caption{16-QAM}
		\label{fig:BER_K_N_b1to3_16QAM}
	\end{subfigure}
	\begin{subfigure}{0.85\linewidth}
		\centering
		\includegraphics[width=\linewidth]{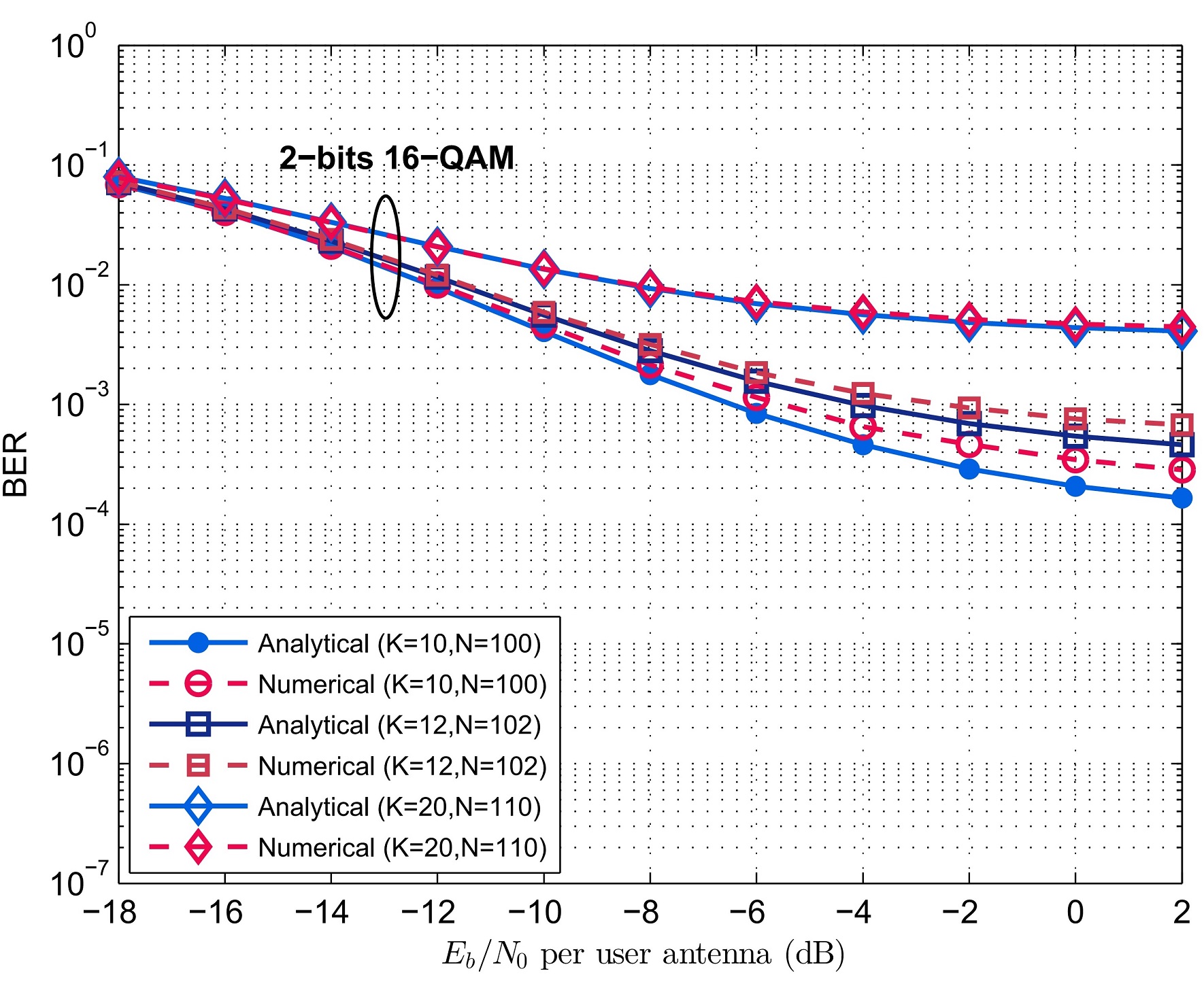}
		\caption{16-QAM}
		\label{fig:BER_K_N_b2_16QAM}
	\end{subfigure}
	\caption{ BER performance of (a) 1-bit QPSK (b) $b$-bits 16-QAM for $b=1$ to 3, and (c) 2-bits 16-QAM, with varying number of  users ($K$), and antennas at the BS ($N$).}
	\label{fig:BER_K_N_QPSK-16QAM}
%		\end{minipage} }
\end{figure}

\subsection{Varying the Number of Users, and BS Antennas}
Authors in \cite{saxena2017analysis} claim that the downlink SER performance of the BS with 1-bit QPSK is the same for any number of BS antennas  ($N$) and users ($K$) that result in the same ratio of $N/K$. We investigated this issue in Figures \ref{fig:BER_K_N_b1_QPSK} and \ref{fig:BER_K_N_b1to3_16QAM} for 1-bit QPSK and (1 to 3)-bits 16-QAM modulations. However, while holding a fixed ratio of $N/K$, we see that results are different and having higher $N$ causes better BER performance for both types of modulations. 
It may happen due to a different definition of SNR  in \cite{saxena2017analysis}. If we consider $\bar{P}$ as the total  transmit power of the users, and replace $\sigma_x^2$ with $\bar{P}/K$ in our equations, we achieve similar results by plotting the SER versus $\bar{P}$. In other words, we observe equal SER performance while having a fixed ratio of  $N/K$. However, this kind of defining the SNR, might be more useful for the downlink scenario.
It is worth noting that both numerical and analytical curves have very close BER results. Therefore, we can find reasons of the BER performance behavior by looking at the analytical expression.

Referring to (\ref{eq:gamma_q0}), the number of users, $K$, appears in the denominator of $\gamma_{q0}$. We further recall that the quantized BER performance is obtained by replacing $\gamma_{0}$ with $\gamma_{q0}$ in (\ref{eq:Bi})-(\ref{eq:BER_MQAM_simple}). Moreover, the expression of the unquantized BER depends on the term $D=N-K$, and it would justify our results in Figures \ref{fig:BER_K_N_b1_QPSK} and \ref{fig:BER_K_N_b1to3_16QAM}. A similar behavior for uplink ZF MIMO is reported in \cite{wang2007performance}.

Furthermore, we have performed another simulation in Fig. \ref{fig:BER_K_N_b2_16QAM} investigating other ways that $K$ and $N$ can be varied. We observe that any equal increase in $K$ and $N$ result a lower BER performance, although a system with higher $N$ has shown higher performance in previous figures. Having a term $K$ in the denominator of $\gamma_{q0}$, might explain such results.
Then, we conclude that if the number of users are increased in the system, we can compensate the performance loss by increasing the number of BS antennas, albeit with more increment for $N$ compared with $K$.

%------------------------------------------------------ Acknowledgements 
%\section{Acknowledgements }
%\label{sec:acknowledgements}

%------------------------------------------------------ Conclusion
\section{Conclusion}
\label{sec:conclusion}
We investigated the effect of coarse quantization on the BER of uplink massive MIMO systems. 
Assuming ZF detection at the BS, we derived a quantized SINR and obtained an analytical BER expression for M-QAM modulations employing low resolution quantizers. 
The proposed expression is a function of quantization resolution in bits. 
The use of uniform and non-uniform quantizers are also investigated numerically, and we found that the analytical expression gives us an upper bound for the BER performance of quantized massive MIMO systems. For the case of non-uniform quantizers,  analytical and numerical BER values are very close, even at very low $b$-bit resolution quantizations of $b=1$ to 3 bits for QPSK, 16-QAM, and 64-QAM modulations. We found that a small BER performance degradation happens for coarse quantized systems of 2-3 bits QPSK and 3-4
bits 16-QAM, compared to the full-precision (unquantized) case. However, increasing the number of BS antennas can compensate the  performance degradation of quantized systems.
We further found that any BER performance loss due to the increase of number of users, $K$, can also be compensated  by increasing the number of BS antennas, albeit with more increment for $N$ compared with $K$.
We generalize our results to include the challenge of channel estimation error at coarse quantized systems, in our future work.

%------------------------------------------------------ Bibliography
%\bibliographystyle{IEEEbib}  % Does not abbreviate the author name
\bibliographystyle{IEEEtran}  % Abbreviate the author nam
\bibliography{ref_twc}        % Read bibitems from "ref_twc.bib" file

\end{document}